# Magnetic and electronic transport percolation in epitaxial GeMn films

N. Pinto, L. Morresi, M. Ficcadenti and R. Murri

*CNR-INFM and Dipartimento di Fisica, Università di Camerino, Via Madonna delle Carceri, I-62032 Camerino, Italy*

F. D'Orazio and F. Lucari

*CNR-INFM and Dipartimento di Fisica, Università di L'Aquila, Via Vetoio-Coppito, I-67010 L'Aquila, Italy*

L. Boarino and G. Amato

*Istituto Elettrotecnico Nazionale "G. Ferraris", Via delle Cacce 91, I-10135 Torino, Italy*

**Abstract**

Electronic transport and magnetic properties of $Ge_{1-x}Mn_x/Ge(100)$ films are investigated as a function of Mn dilution. Depending on *x*, characteristic temperatures separate different regimes in both properties. Resistivity exhibits an insulator-like behavior in the whole temperature range and, below about 80 K, two distinct activation energies are observed. At a higher temperature value, $T_R$, resistivity experiences a sudden reduction. Hall coefficient shows a strong contribution from the anomalous Hall effect and, at $T_R$, a sign inversion, from positive to negative, is recorded. The magnetic properties, inferred from magneto-optical Kerr effect, evidence a progressive decrease of the ferromagnetic long range order as the temperature is raised, with a Curie temperature $T_C$ not far from $T_R$. The transport and magnetic results are qualitatively consistent with a percolation mechanism due to bound magnetic polarons in a GeMn diluted magnetic semiconductor, with localized holes [A. Kaminski and S. Das Sarma, Phys. Rev. B **68**, 235210 (2003)].

I. Introduction

Integration of electron charge and spin degrees of freedom represents the new challenging field of the so called ''spintronics'' or spin polarized electronics [1].
Spin injection into conventional (i.e. non-magnetic) semiconductors (CS) has recently attracted great interest due to the possibility to fabricate new families of spin-dependent electronic devices. In order to inject spin-polarized currents into CS, several scientists have tried to use ferromagnetic (FM) metals as spin sources, forming metal-semiconductor hetero-interfaces. However, the spin polarization of the injected carriers tends to be quickly lost at the hetero-interface via spin-flip scattering due to the dissimilar crystal structure and chemical bonding and to the difference between the charge carriers energy (electrochemical potential) in the ferromagnet and in the semiconductor

[2]. An alternative and promising approach to achieve spin injection into a CS is to use a diluted magnetic semiconductor (DMS), grown introducing a small percentage (< 5 %) of magnetic elements (such as Mn, Co, Ni, and Fe) into the semiconductor host lattice. Recent works have successfully demonstrated electrical spin injection into CS by a DMS spin polarizer [3, 4].

Research activities are mainly focused on Mn-doped compounds of group III-V and II-VI families, where FM films have been grown with a maximum Curie temperature $T_C \approx 110$ K [5]. However, there is considerable interest in higher $T_C$ materials particularly in group-IV elements, owing to their potential compatibility with current Si based technology. Few recent works [6-8] show that Ge crystals doped with transition metal elements do exhibit FM behavior at low temperature. However, the fabrication of good epitaxial quality materials with higher $T_C$ using larger concentration of magnetic elements is inhibited by phase separation of dopant-rich compounds. Moreover, electronic transport properties, such as Hall coefficient and resistivity, have been only partially explored both experimentally and theoretically.

In this work we investigate electronic transport and magnetic properties of thin $Ge_{1-x}Mn_x$/Ge(100) films epitaxially grown at several Mn dilutions in the Ge host. Our transport data evidence that, at a certain temperature $T_R$, the Hall coefficient $R_H$ changes sign and the resistivity, $\rho$, presents a downward kink. The $T_R$ value depends on the Mn concentration and has a strong correlation with the disappearance of the ferromagnetic properties from the films. The temperature behavior of the resistivity below $T_R$ and the magnetic properties can be explained with a percolation model of bound magnetic polarons due to a strong localization of holes.

II. Sample preparation and experimental procedure

Alloys of $Ge_{1-x}Mn_x$ were grown by molecular beam epitaxy (MBE) on epi-ready *n*-type Ge(100) wafers with a nominal resistivity of 2.5 ÷ 7.5 Ω×cm. Ge and Mn were evaporated by standard effusion cells at a constant film deposition rate of ≈ 0.02 nm/s.

Oxide from the Ge wafer surface was thermally desorbed at about 400 °C in presence of an atomic flux of Ge. A clean and oxide free Ge surface, with a perfect 2×1 reconstruction, was observed by in situ x-ray photoelectron spectroscopy (XPS) and reflection high energy electron diffraction (RHEED) analysis, respectively. Prior of GeMn alloy evaporation, a 100 nm thick Ge buffer was epitaxially grown on the wafer. The growth temperature and the thickness of $Ge_{1-x}Mn_x$ films were kept constant at $T_G = 160$°C and 40 nm, respectively. The Mn dilution in the alloy was changed in the range $0.010 \leq x \leq 0.051$. Additional details about growth mechanism, surface morphology and structure of $Ge_{1-x}Mn_x$ films have been reported elsewhere [9].

Hall coefficient and resistivity were measured as a function of the temperature from 18 K to 330 K in a closed cycle He cryostat, having stability better than 0.1 K, by using the van der Pauw configuration. For Hall effect characterization, a DC magnetic field of 850 Oe was used, switched on at low temperature. Surface ohmic contacts were realized by indium, about 5 mm apart, whereas the measurement current was chosen in the range of $10^{-7} \div 10^{-6}$ A.

Magnetic characterization was obtained from a magneto-optical Kerr effect (MOKE) apparatus. The experiments were carried out using radiation incident on the film surface at an angle of 45°. The magnetic field varied in the range ± 5600 Oe and the temperature between 12 K and 310 K. Preliminary measurements were done at the lowest temperature in order to verify the existence of possible wavelength dependence of the MOKE response in the near infrared region. It was observed only a regular variation of the MOKE amplitude, consistent with a frequency dependence of the optical and magneto-optical coefficients. On the contrary, the coercivity and the other features characterizing the hysteresis loop shape remained unchanged. For this reason, the remaining study was carried on at a fixed wavelength (2.00 µm) at which the apparatus gave the highest signal-to-noise ratio. At this wavelength, the penetration depth of the radiation is larger than the thickness of the deposited layer, assuring that the entire MnGe alloy contributes to the MOKE signal. In order to optimize the detected signal we also chose to operate using s-polarized radiation. Differences between longitudinal and polar hysteresis loops are discussed below.

III. Experimental results

a) Hall effect

The Hall coefficient of all investigated $Ge_{1-x}Mn_x$ films decreases of several orders of magnitude as $T$ increases, showing a sign inversion, from positive to negative, at a characteristic temperature $T_R$, whose value (183 K ≤ $T_R$ ≤ 267 K) depends on Mn dilution (Fig. 1). The behavior below $T_R$ is characterized by a dominant contribution from the anomalous Hall effect (AHE) [10, 11], which ceases after the sign inversion. In fact, above $T_R$, the curves are independent on the temperature and perfectly overlap suggesting a strong reduction of any magnetic contribution to the Hall signal and indicating a common origin of the conduction process in our GeMn films. The sign inversion, also observed in GeMn alloys grown at higher $T_G$ [12] and in Mn doped III-V DMS films [13], has been confirmed when the measurements were repeated with different current intensities, whereas it was completely absent in the Ge buffer/Ge(100) system. Unfortunately, the contribution of the AHE to the Hall coefficient does not allow an accurate determination of the charge carrier concentration. Nevertheless, for the whole set of samples we extrapolated a room-temperature hole concentration between ≈$10^{18}$ and ≈$10^{20}$ cm$^{-3}$, from the $R_H$ vs. $T$ behavior (see Table I).

b) Resistivity

The electrical resistivity of $Ge_{1-x}Mn_x$ films is strongly insulator-like in the whole investigated temperature range (Fig. 2), with a downward kink centered at $T_R$ where we observe the sign inversion in $R_H$. At high Mn content, the $\rho$ vs. $T$ curve tends to flatten, in the range 100 K $< T < T_R$, and to reduce the range of resistivity values, indicating a progressive lost of the insulator-like character. Above $T_R$, all the curves perfectly overlap for any composition.

In order to exclude any influence of the substrate to the GeMn transport data, in the inset of Fig. 2 we show the resistance, $R$, as a function of temperature. For comparison, we report also the result for a specifically prepared sample, obtained by deposition of the Ge buffer on the Ge(100) wafer. Below $T_R$, these values are higher than those of the GeMn alloys and with different temperature dependence. However, near $T_R$, we note that the GeMn film resistance drops and tends to assume values and behavior very similar to the Ge substrate (i.e. Ge buffer/Ge(100) wafer). Therefore, above $T_R$, the measured transport properties of the films are strongly influenced by the substrate, in agreement with the conclusions from Hall effect measurements. These results suggest the existence of an energy barrier, below $T_R$, separating the electrical behavior of the film from the substrate.

The temperature dependence of the film resistivity differs from that of a classical highly doped p-type Ge material and does not show any metal to insulator transition (MIT) reported for other DMS systems [13, 14].

Below about 80 K, two thermally activated energies ($E_{a_1}$ and $E_{a_2}$) are found in the film resistivity, as evidenced in the Fig. 3. The measured values of $E_{a_1}$ and $E_{a_2}$ are reported in Table I. Depending on $x$, below 30÷50 K, we measured an activation energy $E_{a_1}$, between 3 and 8 meV. These high values and the corresponding temperature range seem to exclude a variable range hopping conduction mechanism in an impurity band, as reported for doped Ge [8, 15, 16].

At low-intermediate Mn concentration (0.010 ≤ $x$ ≤ 0.033) and in the temperature range 55 K $< T <$ 85 K, we find an activation energy $E_{a_2} \approx 15 \div 20$ meV, which does not correspond to known acceptor energy levels due to Mn doped Ge [17]. For the sample with the highest Mn concentration (5.1%), data analysis, for 18 K $< T <$ 85 K, shows two less distinct values, probably due to a transition towards a different transport mechanism.

The temperature $T_S$, which marks the slope change in the film resistivity curves (see Fig. 3) is reported in Table I, and it will be discussed later.

c) MOKE

In order to look for possible magnetic anisotropy of the sample, we performed MOKE experiments in both a) longitudinal and b) polar geometry, i.e. with the magnetic field a) in the film plane and in the plane of incidence, and b) perpendicular to the film plane, respectively. Figure 4a shows the two

MOKE hysteresis loops at 12 K for the sample with x = 0.051. In the plots, the Kerr signal has been reproduced after subtraction of a linear term, which includes the diamagnetic contributions due to the cryostat windows, the sample substrate and any possible paramagnetic contribution from the film itself. To compare the results obtained in longitudinal and in polar geometry, the vertical scale has been normalized to the highest value for each case. There is evidence that the magnetic anisotropy favors out-of-plane orientation of the magnetization, whereas no in-plane anisotropy is observed when the sample is rotated around an axis perpendicular to the film surface. Moreover, other features, such as the temperature dependence of saturation, remanence, and coercivity are similar for the two (longitudinal and polar) configurations. Analogous results are obtained for the other samples although, due to their lower Mn content, a quantitative comparison is made difficult by the response in longitudinal MOKE which, as typically expected, is one order of magnitude smaller than polar MOKE. Therefore, a detailed temperature dependence study of the magnetic character has been obtained from measurements in the polar geometry only.

Figure 4b shows, as an example, the experimental polar MOKE curves, after subtraction of the linear contribution, relative to the sample with $x = 0.033$ at several temperatures.

The analysis of the hysteresis loops for all samples evidences four temperature ranges. In the first one, starting from the lowest temperature, the film has a FM character with a regular hysteresis loop. The coercivity and the remanence decrease slowly as the temperature is raised. In the second interval, the hysteresis exhibits an irregular shape, which suggests that the sample is magnetically inhomogeneous in this temperature range. This is seen, for example, in the curve at 189 K in Fig. 4. The hysteretic behavior disappears at a characteristic temperature that, for each sample, is comparable with $T_R$ and marks the beginning of the third temperature range, when the sample still shows a magnetic character with a saturating behavior at high fields. This feature persists up to about room temperature, above which, in the fourth temperature range, the film becomes paramagnetic.

In order to synthesize the results for all the samples, in Fig. 5 we show the temperature dependence of the magnetic parameters: a) saturation, b) remanence, and c) coercivity. We notice that the sample with $x = 0.026$ does not follow the general trend of the other samples: its MOKE signal and its coercivity are relatively small. The irregular variation of the magnetic parameters with the temperature suggests the existence of a critical Mn concentration, which could deeply affect the physical properties of the material, as recently suggested for this system [18]. Further experiments will be necessary to clarify this point. For the other samples, the MOKE rotation at saturation is a monotonic function of the concentration and disappears at about 300 K in all films. The remanence and coercivity decrease in a regular manner and become zero at a sample dependent value. We fit the remanence data with the function $(1 - T/T_C)^{1/2}$, which gives an estimate of the Curie temperature $T_C$. Similarly, we observe that the coercivity decreases linearly in a large temperature interval. The fit with the function 1-$T/T_C$ provides an alternative estimate of $T_C$, which is consistent with the value obtained from the remanence. The values of $T_C$ are reported in Table I. Notice that typically there is a residual remanence and coercivity above the calculated $T_C$ (see the insets of Figs. 5b and 5c), in a temperature interval where the remanence shows a slightly positive concavity up to its disappearance at a temperature close to $T_R$.

IV. Discussion

For our GeMn films, the insulator-like electrical behavior, the low hole concentration and the absence of any metal to insulator transition (MIT), are difficult to explain in the framework of standard models adopted in other DMS systems and based on mean field theory [19].

For DMS showing a strongly insulating character, a percolation model has been suggested by Kaminski and Das Sarma to explain both their electrical and magnetic properties [20-22]. The model assumes a heavily compensated material with a random spatial distribution of magnetic impurities. In these systems, charge carriers (i.e. holes) are strongly localized and their spins can form bound magnetic polarons (BMP) with a temperature dependent size. At temperatures of the order of $T_C$ and below it, hole transport in the film occurs by means of nearest-neighbor hopping at localization sites, unoccupied by other localized holes. This model requires that the mean distance between the localized holes (depending on the Mn concentration in the film) must be larger than the hole localization radius, i.e. $a_0^3 p \ll 1$, where $a_0$ represents the characteristic exponential decay of the hole wave function in the localized state and $p$ the effective hole concentration [20].

In this context, it can be proved that the resistivity depends on the temperature following the relation: $\rho \propto \exp(E_{hop}/kT)$, where $E_{hop}$ is a hopping activation energy among localization sites [21]. Just below $T_C$, $E_{hop}$ can be expressed as:

$$E_{hop} = E_{hop}^{(0)} + E_{pol} \quad (\text{for } T < \approx T_C), \quad (1)$$

where $E_{hop}^{(0)}$ is a random energy level mismatch between two localization sites due to the disorder and $E_{pol}$ is the polaron binding energy [21]. At $T \ll T_C$, an infinite BMP cluster will span the whole film, and a hole will jump in a localization position already polarized. In this case, the impurity polarization will not give any contribution to the hopping activation energy:

$$E_{hop} = E_{hop}^{(0)}. \quad (2)$$

For our GeMn films, we have estimated the $a_0$ value considering a hydrogenic model, for which $a_0 = \hbar/\sqrt{2\varepsilon E_b} \simeq 2 \times 10^{-9}$ m [23] using $\varepsilon = 16.2$ as Ge dielectric constant [17] and $E_b = 30$ meV [24] as hole binding energy. Taking the $p$ values reported in Table I, we get $8 \times 10^{-3} \leq a_0^3 p \leq 8 \times 10^{-2}$. Though these values are slightly higher than those assumed by the model, some general and interesting considerations can be done [21]. In fact, considering the applicability of the percolation model to our system, $E_{hop}$ must be identified with the two activation energies $E_{a_1}$ and $E_{a_2}$ extrapolated from the resistivity curves, at $T<<T_C$ and $T<\approx T_C$, respectively. Their difference gives the polaron energy, of the same order of magnitude of those reported for other DMS systems [25]. We note that $E_{hop}^{(0)}$ (i.e. $E_{a_1}$) and $E_{pol}$ (i.e. $E_{a_2} - E_{a_1}$) show two opposite dependences on the Mn content (see Table I). As $x$ rises up to $\approx 0.03$, $E_{hop}^{(0)}$ decreases as expected by a reduction of the disorder in the alloy, while $E_{pol}$ rises since it linearly depends on $x$ [21]. At higher Mn concentration, the behavior of the energy suggests a change of the transport mechanism. In fact, for $x = 0.051$, the quantity $a_0^3 p$ is probably too high ($\approx 0.1$) to assume hole localization and it may represent a transition value approaching the case of charge carrier localization inside a cluster ($a_0^3 p \gg \approx 1$) [21]. This cluster is formed by a grouping of magnetic impurities, probably due to local inhomogeneities in the Mn distribution [21], as suggested by MOKE data. Although in this case the model predicts the possibility of a change in the monotonic behavior of the resistivity (a maximum in the curve), its absence may be attributed to a relatively small exchange interaction between holes (hopping between clusters) and magnetic impurities inside the clusters, with respect to the Coulomb repulsive interaction between holes within a cluster (see Eq. 18 in Ref. 21). This fact could be responsible for the two similar activation energies ($E_{a_1} \cong E_{a_2}$) found for $x = 0.051$.

For the other three samples (with $x \leq 0.033$), the characteristic temperature, $T_{cover}$, at which the infinite cluster covers most of the sample and the resistivity curve changes its slope, is given by [20]:

$$T_{cover} \approx T_C \exp\left(-0.86 / (a_0^3 p)^{1/3}\right) \qquad (3)$$

We tentatively tried to estimate the $T_{cover}$, applying the relation (3) to our films by using the data quoted in Table I. The resulting values as a function of the Mn concentration are reproduced in Fig. 6. For comparison, this figure shows also the temperature, $T_S$, corresponding to the slope change in the resistivity, as reported in the Fig. 3. We note that the two curves in Fig. 6 present a qualitatively similar behavior (i.e. they are almost parallel), suggesting a possible strong relation between $T_{cover}$ and $T_S$. In particular, the absolute difference between $T_{cover}$ (obtained from the relation (3)) and $T_S$ is largely due to the well known underestimation of charge carrier density in DMS systems, due to the presence of AHE [6]. This result suggests that the percolative conduction model can qualitatively explain the resistivity behavior of our GeMn films. We remark that, for the sample with the highest Mn content ($x = 0.051$), a percolative model can be still invoked, but with holes localized inside the clusters [21].

The percolative model can also explain the magnetic properties of our GeMn films, although the temperature dependent interconnection of the magnetic regions influences differently the magnetic and the transport responses. For example, the irregular shape of the MOKE loops observed at intermediate temperatures may be due to an additional contribution from weakly interacting magnetic clusters which do not have a cooperative magnetic response and are not included in an infinite electronic transport path. We note that the temperature dependence of the remnant magnetization (Fig. 5b) results to be convex in a large interval of temperatures, contrarily to what predicted from the percolation theory [21]. This may be related to the high value of $a_0^3 p$ and, therefore, to departure from the low concentration limit of magnetic impurities and charge carriers. It would also explain the relatively high values of $T_C$. Around this temperature, as the magnetic interaction is weakened, the concave behavior is recovered (Fig. 5b, inset).

At the moment, we can not exclude that this inhomogeneous phase could be made of small precipitates (typically $Mn_{11}Ge_8$ and $Mn_5Ge_3$) that can plague the growth of the GeMn films [26], although our previous diffraction analysis did not evidence them [9]. Moreover, the particular growth conditions of our GeMn alloys and their magnetic parameters ($T_C$, etc.) seem to exclude that these precipitates, if there are any, belong to one of the known Mn-rich GeMn phases [27]. However, their eventual presence in the film does not change the general conclusion derived from the applicability of the present percolation model to the transport and magnetic properties of the DMS phase.

There is a final intriguing point to be discussed: the complete disappearance of the FM character at $T_R$, associated to the sign inversion of charge carriers and to the abrupt drop of the film resistivity. For our films, the low thickness and the presence of an island morphology cause an high scattering of charge carriers and then an higher resistance compared to the Ge substrate (see the inset in Fig. 2). These facts suggest the presence of an energy barrier, below $T_R$, between the *p*-type doped film and the *n*-type doped Ge wafer that prevents appreciable phenomena of parallel conductance through the Ge substrate. The formation of a depletion layer, spatially extended in the Ge wafer, enhances the electrical separation between the two systems. Therefore, below $T_R$, the transport measurements of GeMn alloys are negligibly influenced by the substrate. However, above $T_R$, according to our interpretation, the thermal energy would overcome this barrier allowing electrons to contribute to the charge transport of the whole structure, with consequent drop of the resistivity and sign change of the Hall coefficient. This mechanism would be also responsible for electron-hole recombination. The consequent strong reduction of holes, responsible for the long range exchange interaction between magnetic ions, could explain the disappearing of the magnetic character of the samples [20-22].

## V. Conclusions

The magnetic and transport properties observed in our GeMn films are consistent with a BMP percolation model, applied to the diluted phase. In fact, starting from low temperature, an infinite percolative polaron cluster, formed merging single clusters of BMP, spans the whole sample up to $T_S$, establishing also a ferromagnetic correlation across the whole sample. Rising the temperature above $T_S$, the infinite cluster disappears and finite size regions (clusters of BMP, still interacting ferromagnetically) are responsible for the regular decrease of the remanence and the coercivity as the temperature increases. At $T_C > T_S$, polarons decrease their size ($r_{pol}(T) \propto 1/T$) below a critical value which prevents zero-field long-range exchange interaction. Finally, at $T_R > T_C$, we observe a resistivity drop and sign inversion of charge carriers due to a prevailing transport contribution from electrons of the Ge substrate. The appearance of a transport mechanism dominated by electrons could explain the vanishing of any ferromagnetic feature in our GeMn films.

**Table I**

Electronic and magnetic parameters of $Ge_{1-x}Mn_x$ films epitaxially grown at 160°C. From left to right the columns indicate: sample name; composition; hole concentration from extrapolation to R.T. of the Hall coefficient (see the text); activation energies in two different temperature ranges below about 80 K; $T_S$ and $T_{cover}$ respectively (see the text); Curie temperature from MOKE data; sign inversion temperature ($T_R$) of the Hall coefficient.

| Sample | x | $p$ (cm$^{-3}$) | $E_{a1}$ (meV) | $E_{a2}$ (meV) | $T_S$ (K) | $T_{cover}$ (K) | $T_C$ (K) | $T_R$ (K) |
|---|---|---|---|---|---|---|---|---|
| Mn46 | 0.010 | $1 \times 10^{18}$ | 8 ± 0.1 | 15.5 ± 0.2 | 36 | 3 | 185 | 267 ± 3 |
| Mn51 | 0.026 | $3 \times 10^{18}$ | 3.1 ± 0.1 | 15.5 ± 0.2 | 45 | 7 | 140 | 183 ± 3 |
| Mn41 | 0.033 | $5 \times 10^{18}$ | 3.1 ± 0.1 | 20.5 ± 0.5 | 50 | 16 | 200 | 250 ± 3 |
| Mn44 | 0.051 | $1 \times 10^{19}$ | 6.1 ± 0.2 | - | - | 32 | 235 | 248 ± 3 |

**Figure captions**

**Figure 1**
Absolute Hall coefficient values measured at 850 Oe as a function of the temperature for the investigated set of GeMn alloy films. Squares: $Ge_{0.99}Mn_{0.01}$; triangles: $Ge_{0.949}Mn_{0.051}$. Open and closed symbols indicate positive and negative values, respectively.

**Figure 2**

Resistivity as a function of temperature. Squares: $Ge_{0.99}Mn_{0.01}$; open triangles: $Ge_{0.949}Mn_{0.051}$. In the inset, the resistance as a function of temperature is shown, including also the data relative to the sample constituted only by the Ge buffer on the Ge(100) wafer (diamonds).

**Figure 3**

Arrhenius plot of the resistivity vs. the reciprocal of the thermal energy. Squares: $Ge_{0.99}Mn_{0.01}$; crosses: $Ge_{0.974}Mn_{0.026}$; circles: $Ge_{0.967}Mn_{0.033}$; open triangles: $Ge_{0.949}Mn_{0.051}$. The curves change their slope at a characteristic temperature, $T_S$. The arrow indicates $T_S$ for the $Ge_{0.967}Mn_{0.033}$ film.

**Figure 4**

a) Longitudinal (•) and polar (□) MOKE hysteresis loops for the $Ge_{0.949}Mn_{0.051}$ film at 12 K. The vertical scale has been renormalized to allow a comparison. The saturation MOKE rotation is 0.0018° and 0.0423° for longitudinal and polar geometry, respectively. The difference between these values causes the evident smaller signal-to-noise ratio for the longitudinal hysteresis. b) MOKE hysteresis loops for the $Ge_{0.967}Mn_{0.033}$ film at several temperatures. Notice the irregular shape of the loops at intermediate temperature (189 K) suggesting an inhomogeneous magnetic structure.

**Figure 5**

Temperature dependence of the magnetic parameters measured by MOKE for the $Ge_{1-x}Mn_x$ films: a) saturation; b) remanence; c) coercive field. The symbols refer to the different values of $x$, as explained in the legend of b). The lines in a) are a guide to the eye; in b) and c) are the fitting curves following a relation of the form $(1-T/T_C)^\beta$ with $\beta = 1/2$ and $\beta = 1$ for b) and c), respectively. The insets in b) and c) show in detail the data close to $T_C$; the axis labels are the same as for the main figures and the lines are guides to the eye. Note that the inset b) reveals a concavity of opposite sign with respect to the lower temperature trend for remanence.

**Figure 6**

Dependence of $T_{cover}$ (triangles) and $T_S$ (circles, see the inset of Fig. 2) on Mn content, $x$. The values of $T_{cover}$ have been computed using Eq. (3) and the data quoted in Table I, with $a_0 = 2 \times 10^{-9}$ m. The two lines are linear regressions of the data points.

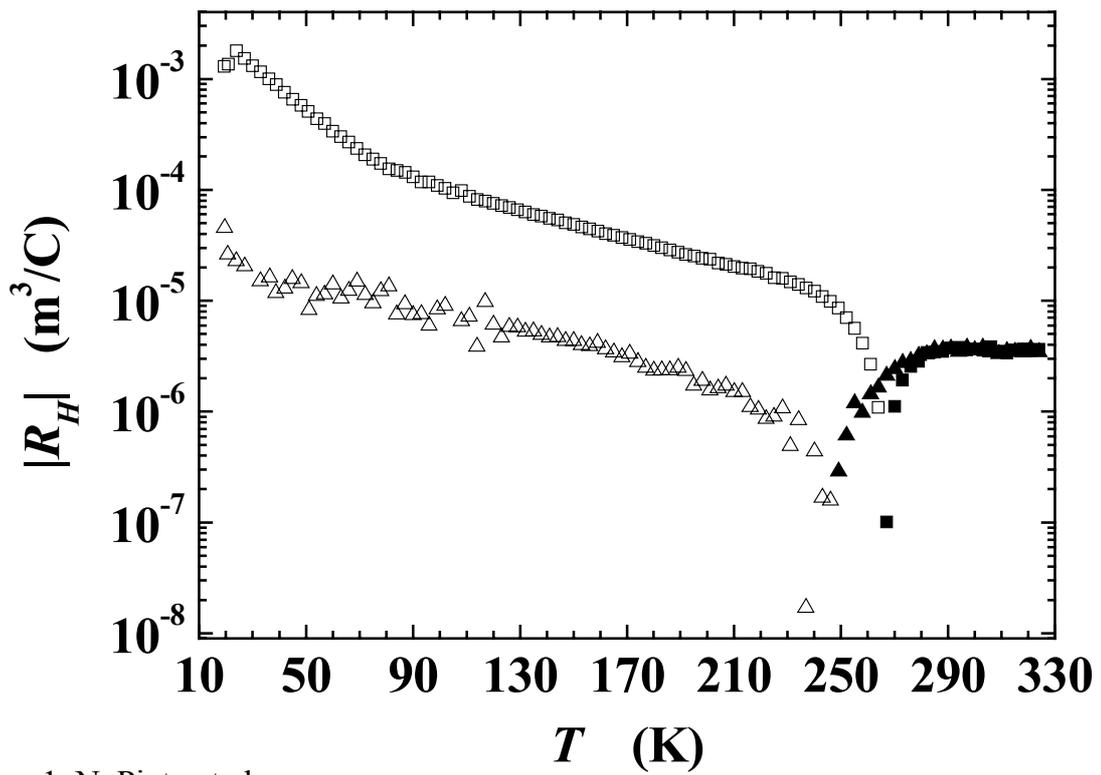

Figure 1, N. Pinto et al.,

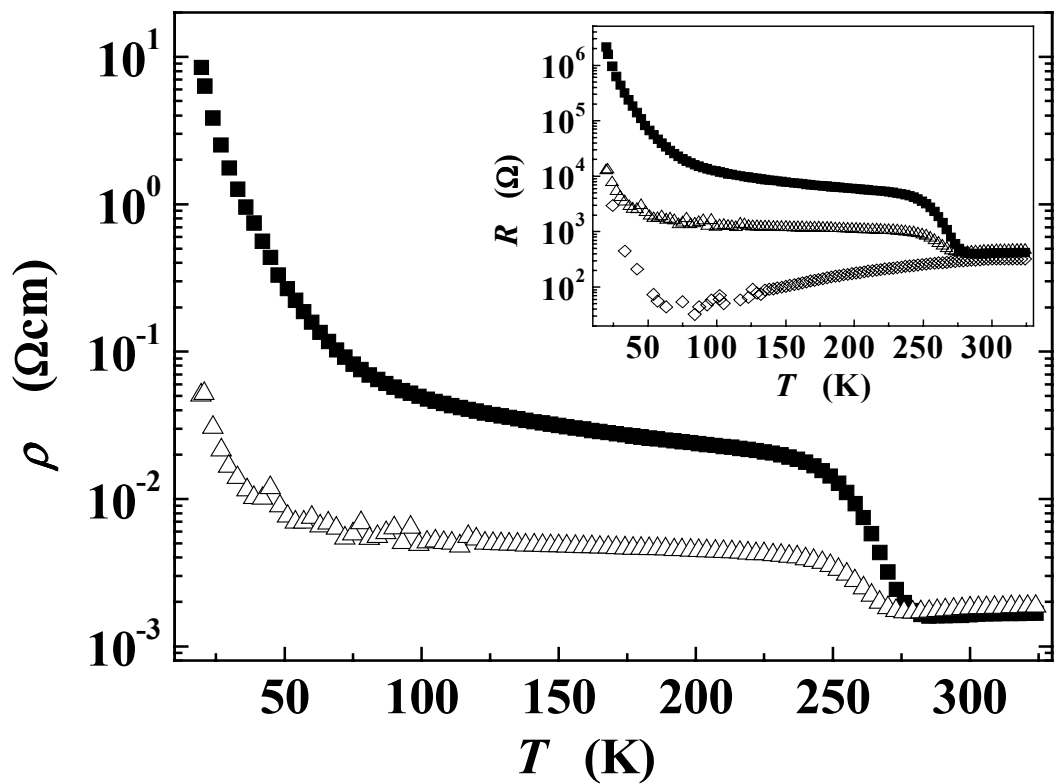

Figure 2

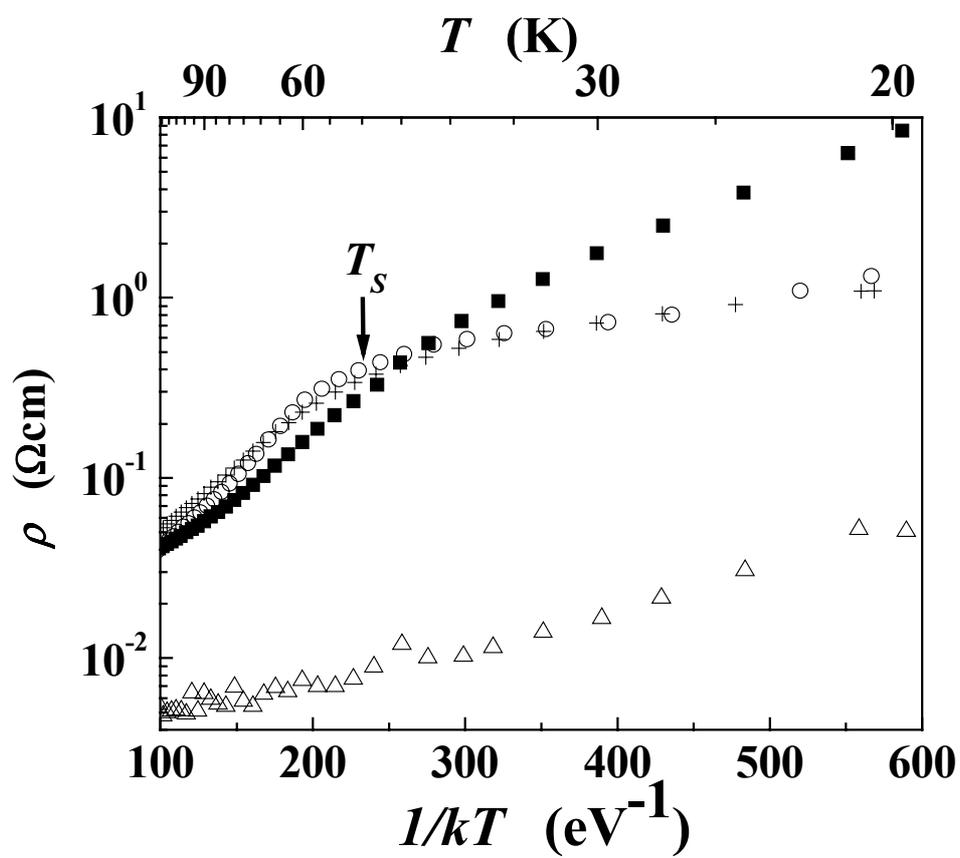

Figure 3

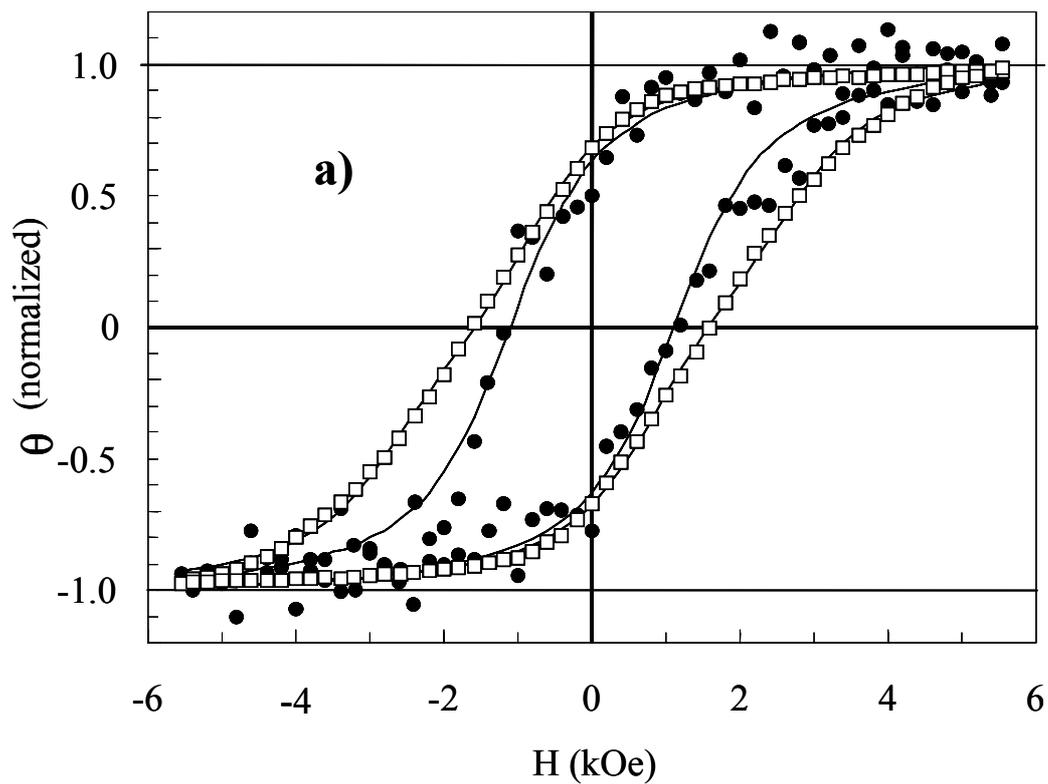
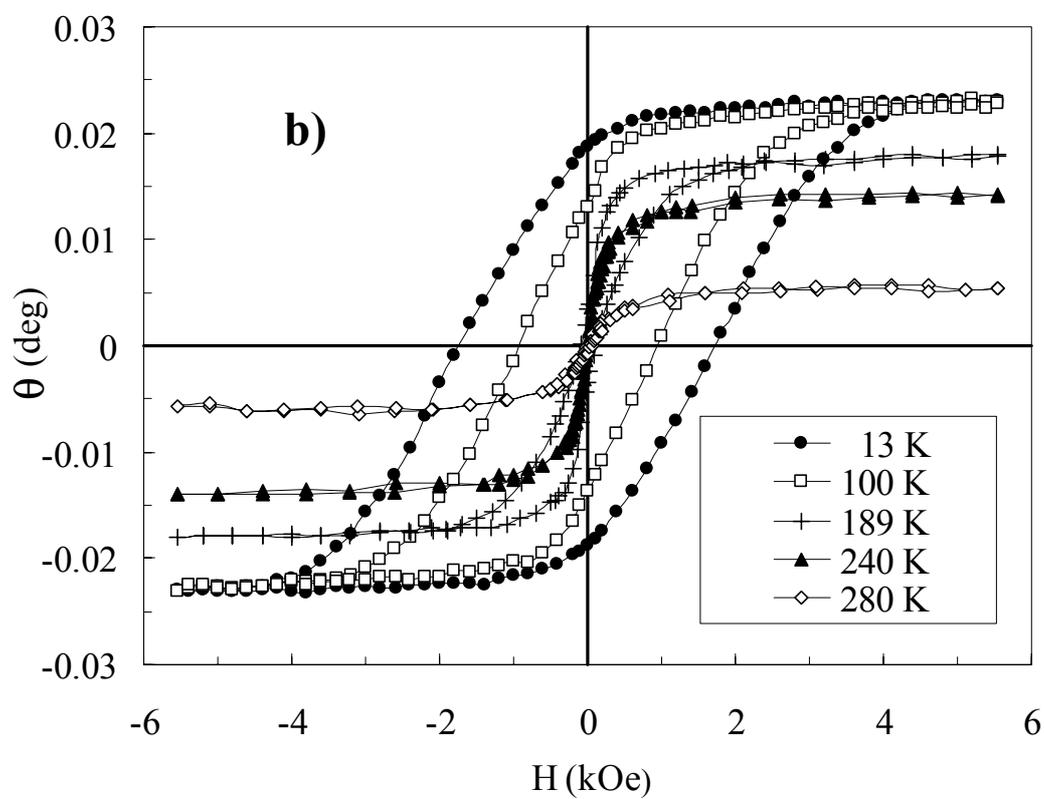

Fig. 4

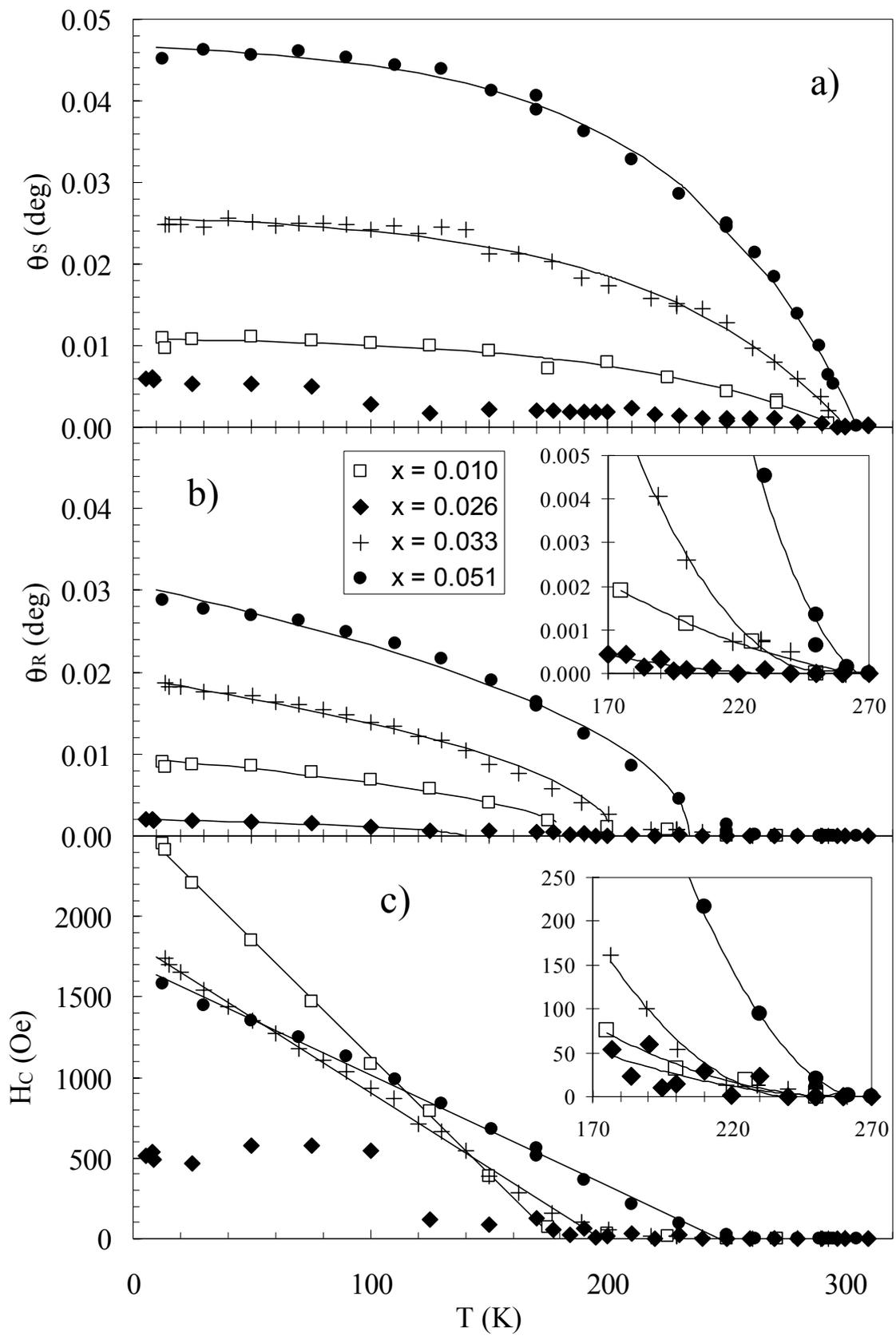

Fig. 5

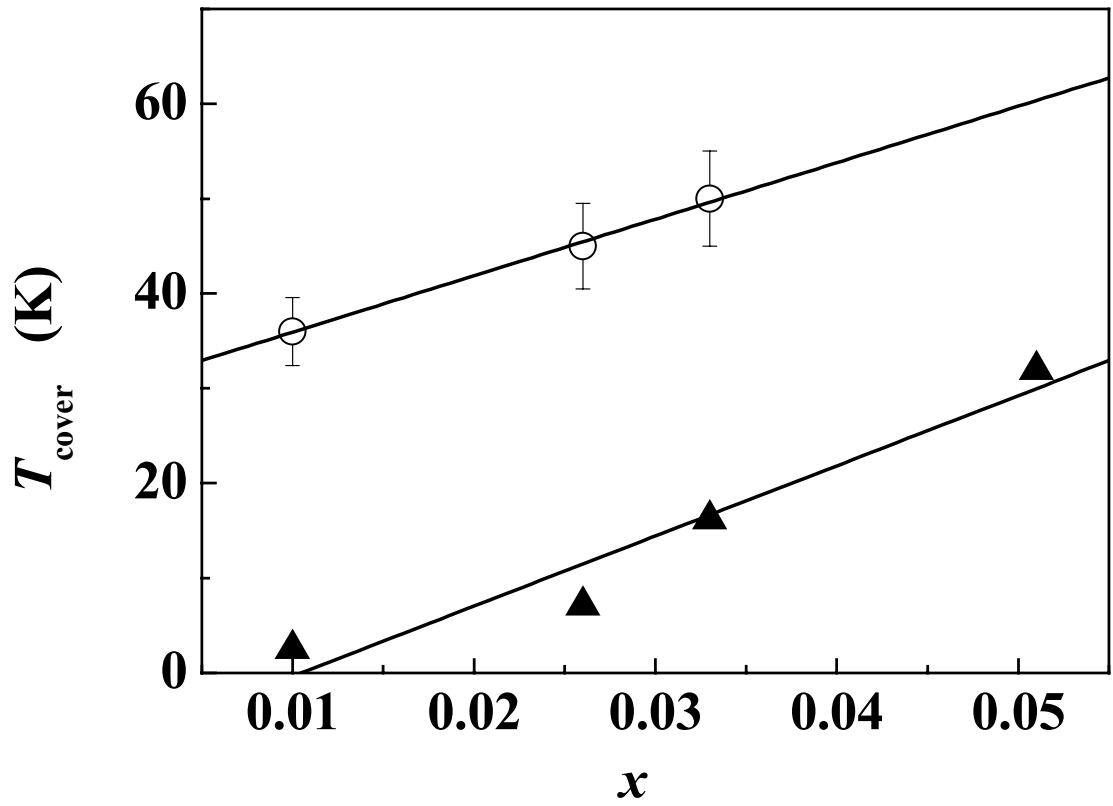

Fig. 6